\documentclass[english,aps,preprint]{revtex4-1}
\usepackage[T1]{fontenc}
\usepackage[latin9]{inputenc}
\usepackage{color}
\usepackage{amsmath}
\usepackage{amssymb}
\usepackage{graphicx}
\usepackage{physics} % add bra and ket 
\PassOptionsToPackage{normalem}{ulem}
\usepackage{ulem}

\makeatletter
\@ifundefined{textcolor}{}
{%
 \definecolor{BLACK}{gray}{0}
 \definecolor{WHITE}{gray}{1}
 \definecolor{RED}{rgb}{1,0,0}
 \definecolor{GREEN}{rgb}{0,1,0}
 \definecolor{BLUE}{rgb}{0,0,1}
 \definecolor{CYAN}{cmyk}{1,0,0,0}
 \definecolor{MAGENTA}{cmyk}{0,1,0,0}
 \definecolor{YELLOW}{cmyk}{0,0,1,0}
}

\usepackage{braket}
\usepackage{bbold}

\makeatother

\usepackage{babel}
\begin{document}

\title{\LARGE{Transport in quantum chains under strong monitoring}}

\author{\large D. Bernard}

\author{\large T. Jin}
%\email{jin@lpt.ens.fr}

\author{\large O. Shpielberg}
%\email{ohad19@gmail.com}

%\author{\large D. Bernard - have we forsaken alphabetical ordering ?}
%\email{denis.bernard@ens.fr}

\selectlanguage{english}%

\affiliation{Laboratoire de Physique Th\'eorique de l'\'Ecole Normale Sup\'erieure de Paris, CNRS, ENS, PSL University \& Sorbonne Universit\'e, France. \\~}

\begin{abstract}

We study the transport properties of quantum $1D$ systems under strong monitoring. The quantum Zeno effect inhibits transport and induces localization. Beyond the Zeno freezing and on long time scales, a new  dynamics  emerges in the form of a Markov process. Studying fermionic and bosonic chains under strong monitoring, we are able to identify the quantum origin of the classical exclusion process, inclusion process and a sub-class of the misanthrope process. Moreover, we show that passive monitoring cannot break time-reversal symmetry and that the transport generally loses its ballistic nature existing for weak measurements. 

\end{abstract}

\maketitle

\section{{\small{}introduction}}

Recently, the study of many-body quantum systems has taken a prominent role due to new horizons of experimental plausibility, especially using ultra cold gases, see \cite{dalibard-rev-mod-phys, cazalilla-et-al} and reference therein. The scope of ``many-body quantum systems'' is wide and  includes isolated systems \cite{polkonivkov-et-al}, quantum quenches \cite{,calabrese-cardy,revue-quench},  a coupling to a reservoir \cite{Prosen2011} and many more \cite{nature-phys-revue} (and references therein). In this study, we wish to explore some aspects of many-body quantum systems  and of transport dynamics under continuous and strong measurements.  

When dealing with quantum transport, a fundamental question that arises is how the system dynamics is altered by the measurement process and what are the consequences on the transport properties. This is for instance of particular relevance for ultracold atomic and molecular gases in optical lattices. 
Intuitively, we expect the emerging transport dynamics to not only  be induced by the measurement but to also  depend on the monitoring process. This effect has been seen experimentally in \cite{key4} and discussed theoretically in \cite{key3,key3_1}.

Famously, a system under continuous and strong measurement is repeatedly projected onto a single eigenstate. Effectively, the system  persists at that state for a long time. This is known as the quantum Zeno effects~\cite{zeno}. However, at long enough time scales, the system can jump between the eigenstates of the measurement, introducing a rich dynamics at a slow time scale.  Here, we report on the dynamics of some simple quantum chains (fermionic and bosonic), under continuous measurement. 

A direct consequence of the Zeno effect is to inhibit transport. Even if the many-body system is driven out-of-equilibrium, the asymptotic system steady state supports neither currents nor long range correlations and hence reflects some kind of induced localization~\cite{local1,local2,many-local,many-local2}. Indeed, besides projecting the system state onto eigenstates, quantum monitoring back-action induces random stochasticity into the system dynamics which destroys coherences and produces localization. This is analogous to the observed many-body localization induced by stochastic randomness in models of critical systems~\cite{Bernard-Doyon-17}.

It was previously noticed that for strong and continuous measurement, the slow dynamics between the measurement eigenstates forms a Markov  process \cite{key2,key2_2,BBT-15,Frohlich-et-al}. It is interesting to check whether  the quantum origins of some canonical Markov processes can be found. Here, we will do just that. We will show which quantum setup leads to the symmetric exclusion process (SSEP), the inclusion process and a sub-class of the misanthrope processes. The fact that some of these processes are exactly solvable allow us to decipher statistical properties of quantum transport under strong monitoring.

Because the monitoring apparatus act as macro- or meso-scopic devices interacting with the many-body system and hence induce dissipation, these slow dynamics are locally diffusive -- or at least sub-ballistic --  even if the transport in the un-monitored system is ballistic. These dynamics are classical because strongly monitoring a system projects the latter on the measurement eigenstates, called pointer states. The emergent classical dynamics is therefore dependent on the monitoring process. Nevertheless, echoes of the quantum origin of these classical dynamics remain.

The method to find the emerging Markovian dynamics is general and can be used in related experiments for consistency checks in the study of quantum transport processes with measurements, or for quantum systems continuously interacting with a reservoir.  

The outline of the paper is the following. In Sec.~\ref{sec:RIP and sto Lind} we recap the repeated interaction technique that produces quantum trajectories and discuss the limit of strong measurement. In Sec.~\ref{sec:Results} we present the emerging dynamics of the XY spin chain under strong measurement of the local $\sigma^z$. Moreover, we discuss the emerging dynamics of the bosonic tight-binding Hamiltonian under strong measurement of the local occupancy as well as generalizing to variants of the local occupancy. Finally, a detailed discussion is performed in  Sec.~\ref{sec:Discussion}, suggesting possible implications to the obtained results. 

\section{The repeated measurement procedure and quantum monitoring
\label{sec:RIP and sto Lind}}

In this section, we recall how the repeated measurement procedure produces a stochastic Lindblad equation, whose trajectories describe the evolution of a monitored quantum system ~\cite{book-qmeasure,book-qmeasure_2}. These are the so-called quantum trajectories~\cite{qtraj-hist,qtraj-hist_2,qtraj-hist_3}. It will serves us as the starting ground for the processes we wish to consider. Then, we discuss the effective dynamics emerging at the strong measurement limit. 

\subsection{Quantum trajectories}

Let us consider a quantum system with density matrix $\rho$ and a series of probes, all prepared in the state $\ket{\varphi}\bra{\varphi}$. A single probe is sent to interact with the system for a short time, after which it is measured with respect to some observable. This procedure is repeated indefinitely. 
Suppose that $s$ and $\ket{s}\bra{s}$ form the complete set of eigenvalues and projectors of the probe measurement. The evolution of the system after interacting and measuring the probe at state $s$ is given by $\rho \rightarrow \frac{F_s \rho F^\dagger _s}{\pi(s)} $, where the Kraus operators are $F_s = \bra{s} U \ket{\varphi}$ and $\pi(s) = \Tr ( F_s \rho F^\dagger _s) $ are the associated probabilities to measure the probe in state $\ket{s}$. Here $U$ is the unitary evolution operator for the system + probe dynamics prior to the measurement. Notice that $F_s$ acts on the space of the system only, tracing out the probe's degrees of freedom. The unitarity of $U$ implies that the Kraus operators satisfy $\sum_s F^\dagger _s F_s = \mathbb{1} $ and this ensures that the conservation of probability $\sum_s \pi(s) =1$ is maintained.

Now, let us assume a continuous evolution of the system's density matrix $\rho$. Namely, we consider each ``turn'' of interaction + measurement with a probe to happen at a short time $dt$ and to slowly change the density matrix at the $dt$ scale. 

A simple method to achieve that is to require the Kraus operators to be composed of normalized unity operators with perturbative expansions scaling with  $\sqrt{dt}$, see e.g.~\cite{book-qmeasure,book-qmeasure_2,yann-steph,Pellegrini}. The continuous evolution of the density matrix is captured by stochastic Lindblad equations, called quantum trajectory equations \cite{qtraj-hist,qtraj-hist_2,qtraj-hist_3},
\begin{equation}
d\rho_t = -\frac{i}{\hbar}\left[H,\rho_t \right]dt + \eta \nu_f L_N(\rho_t)dt + \sqrt{\eta \nu_f}\, M_N(\rho_t)\, dB_t.  \label{eq:Sto Lind} , 
\end{equation}
with $L_N(\rho) = N\rho N^\dagger - \frac{1}{2} \lbrace N^\dagger N ,\rho \rbrace$ and $M_N(\rho) = N\rho + \rho  N^\dagger - \rho \Tr ( N\rho + \rho  N^\dagger)$. 
 Here $H$ is the Hamiltonian of the system and $N$ is an operator associated with the interaction and measurement of the probes~\footnote{Eq.\eqref{eq:Sto Lind} can be generalized to include a series of different measurement operators $N_i$ accompanied by their associated Brownian motions $B^i_t$.}. The $\left[\cdot,\cdot\right]$ and $\lbrace \cdot,\cdot\rbrace$ are the standard commutation and anti-commutation operators and $dB_t$ is the standard It\^o increment satisfying $dB^2_t= dt$.  The cumulated classical random signal $S_t$ produced by the monitoring process changes in time according to $dS_t=\eta \nu_f \Tr ( \rho_t(N +  N^\dagger))\, dt + dB_t$. Its time drift is governed by the time dependent expectation of the system observable $N+N^\dag$ and hence provides a continuous monitoring of that observable. 
 Furthermore, $\eta \nu_f$ determines the rate at which information is extracted and $\eta$ is a dimensionless parameter we will vary in what follows. 
 
Discarding the outcomes of the measurements leads to a mean dynamics, i.e. averaging with respect to the possible quantum trajectories of $\rho$. This  yields the (mean) Lindblad evolution equation
\begin{equation}
\frac{d}{dt}\bar{\rho}_t = -\frac{i}{\hbar}\left[H,\bar{\rho}_t \right] + \eta \nu_f L_N(\bar{\rho}_t). \label{eq:mea Lindblad}
\end{equation}
 The presence of the second Lindblad term reflects the dissipation induced by the measurement back-action.

The Lindblad equation is the most general Markovian evolution equation which is trace preserving and completely positive. It can also be derived from different settings \cite{Zoller_Book,Breuer_Book,Lindblad1976}. However, we consider here the  repeated interaction procedure to allow putting the results of the paper in a concrete  experimental context. 

In what follows, we consider a $1D$ lattice, with a series of localized measurement operators $N_i=N_i^\dag$, with $i$ indexing the lattice site, that have a non-degenerate spectrum~\footnote{We here restrict ourselves to self-adjoint measurement operators, $N=N^\dag$. Measurement with non-self adjoint operators may also be considered.}. Thus we may rewrite the mean Lindblad equation in the form 
\begin{equation}
\frac{d}{dt}\bar{\rho}_t = L(\bar{\rho}_t) + \eta\, L_b(\bar{\rho}_t) \label{eq:Lindblad on a lattice}
\end{equation}
with $L(\rho) = -\frac{i}{\hbar}\left[H,\rho \right] $ and $L_b(\rho) =  \nu_f \sum_j L_{N_j}(\rho)$. The system Hamiltonian $H=\sum_i h_i$ is the sum of local interactions.

\subsection{Strong measurements}
Let us consider the limit $\eta \rightarrow \infty$, where the dissipative part of the Lindblad equation (\ref{eq:Lindblad on a lattice}) dominates the Hamiltonian evolution. Here, any density matrix not belonging to the kernel of $L_b$ is exponentially suppressed in time. Hence, the system is projected to a state in the kernel $\ker L_b$ and remains there for a long time. This is a manifestation of the quantum Zeno effect. 
At the time scale $s= t/\eta$ for $t,\eta \rightarrow \infty$ a non-trivial dynamics emerges, where the density matrix exhibits an interplay between the pointer states, i.e. the eigenstates of the measurement operators which form a basis of $\ker L_b$.

The simplest approach to capture the emerging dynamics of strong dissipation consists of looking at the mean evolution (\ref{eq:Lindblad on a lattice}). Second order perturbation theory then yields that the mean effective evolution equation is $\frac{d}{ds}\bar{\rho}_s  =  \mathcal{L}(\bar{\rho}_s )
$ where 
\begin{equation}
\mathcal{L}(\rho ) = - \Pi_0 L (L^\perp _b)^{-1}  L \Pi_0\, ( \rho).
\label{eq: strong diss evo}  
\end{equation}
Here, $\Pi_0$ is the projector onto $\ker L_b$ and  $(L^\perp _b)^{-1}$ denotes the inverse of the restriction of $L_b $ onto the complement of $\ker L_b$.  Since the slow dynamics is composed of the interplay between the pointer states, the mean density matrix can be written as
\begin{equation}
\bar{\rho}_s = \sum_{{\epsilon} } \bar Q_s({\epsilon}) \mathbb{P}({\epsilon}),
\end{equation}
where $\mathbb{P}_{{\epsilon}}$ are the projectors onto the pointer states denoted by $|\epsilon\rangle$ and $\bar Q_s(\epsilon)$ are their respective time-dependent weights, with $\sum_\epsilon \bar Q_s(\epsilon)=1$. Therefore, the evolution of the mean density matrix $\bar{\rho}_s$ is  contained in the time evolution of the $\bar Q_s(\epsilon)$. 

A more informative approach to the effective slow dynamics consists of looking at the quantum trajectories for the system density matrix $\rho$, whose evolutions are governed by the stochastic evolution equation (\ref{eq:Sto Lind}), in the limit $\eta\to\infty$ at fixed $s=t/\eta$. One then learn \cite{BBT-15,Frohlich-et-al} that, at any given fixed time $s$, the system is in one of the pointer states, with probability one. So that, at any fixed time $s$, the system is in a pure, but random, time dependent pointer state $\rho_s=\mathbb{P}({\epsilon_s})$. The probability for the system to be in a given pointer state $\mathbb{P}({\epsilon})$ is $\bar Q_s(\epsilon)$. As shown in \cite{BBT-15,Frohlich-et-al}, the slow dynamics is then reduced to Markovian quantum jumps from one pointer state to another with probability rate depending on the system Hamiltonian and measurement operators~\footnote{The convergence of the quantum trajectory dynamics to a Markov chain on the set of pointer states is weak in the sense it only claims the convergence of the N-point functions. It is not a strong convergence.}. The Markovian evolution of the probabilities $\bar Q_s(\epsilon)$ is then equivalent to (\ref{eq: strong diss evo}).

For our purpose, it is sufficient to say that $\bar Q_s(\epsilon)$ follows a Markovian evolution. In what follows we will identify the emerging Markovian dynamics for a few choice of many-body Hamiltonian dynamics.

\section{Results
\label{sec:Results}}
In this section we derive our  main results. Namely, we find the Markovian dynamics describing the large $\eta$ limit of a few choice Hamiltonians and measurements.

\subsection{Spin chain with strong local $\sigma^z$ measurements
\label{subsec: spin chain }}

Let us consider a periodic system of $ L$ sites occupied by spin $\frac{1}{2}$ fermions. The system evolves according to the XY Hamiltonian $H= \varepsilon \sum_j (\sigma^x _j \sigma^x _{j+1} + \sigma^y _j \sigma^y _{j+1}) $ and the measurement operators are $N_j = \sigma_j ^z$. The mean dynamics (\ref{eq:Lindblad on a lattice}) is then that of the XY model with dephasing noise, see e.g. \cite{dephasing,dephasing2,dephasing3}, but the stochastic quantum trajectories are typically different from this mean evolution. 

At the large $\eta$ limit, using \eqref{eq: strong diss evo} we find 
\begin{equation}
\frac{d}{ds} \bar{\rho}_s = -\frac{1}{2}D \sum_j \left[ \sigma^+ _j \sigma^- _{j+1}, \left[  \sigma^- _j \sigma^+ _{j+1}, \bar{\rho}_s  \right]   \right] + \textnormal{h.c.}, \label{eq: eff dyn spin chain}
\end{equation}
where $\sigma^\pm  = \sigma^x \pm i\sigma^y   $ and $D= \frac{\varepsilon^2}{\hbar^2 \nu_f}$. 

The pointer states here are $\mathbb{P}(\epsilon) = \otimes_j\, \mathbb{P}^{\varepsilon_j} _j$, where $\varepsilon_j = \pm$ and $\mathbb{P}^{\varepsilon_j} _j$ are the projectors onto the local $\ket{\pm}$ states, so that strongly measuring the local $\sigma^z$ gives access to the instantaneous spin profile. Using the local two-site pointer states, one  obtains 
\begin{eqnarray}
\frac{1}{D}\mathcal{L}(\mathbb{P}^{+} _j \otimes \mathbb{P}^{-} _{j+1}) &=& \mathbb{P}^{-} _j \otimes \mathbb{P}^{+} _{j+1}-\mathbb{P}^{+} _j \otimes \mathbb{P}^{-} _{j+1} \nonumber
\\ 
\frac{1}{D}\mathcal{L}(\mathbb{P}^{-} _j \otimes \mathbb{P}^{+} _{j+1}) &=& \mathbb{P}^{+} _j \otimes \mathbb{P}^{-} _{j+1}-\mathbb{P}^{-} _j \otimes \mathbb{P}^{+} _{j+1}
\\ \nonumber
\mathcal{L}(\mathbb{P}^{+} _j \otimes \mathbb{P}^{+} _{j+1}) &=& 0
\\ \nonumber
\mathcal{L}(\mathbb{P}^{-} _j \otimes \mathbb{P}^{-} _{j+1}) &=& 0.
\end{eqnarray}

\begin{figure}
\begin{centering}
\includegraphics[scale=0.25]{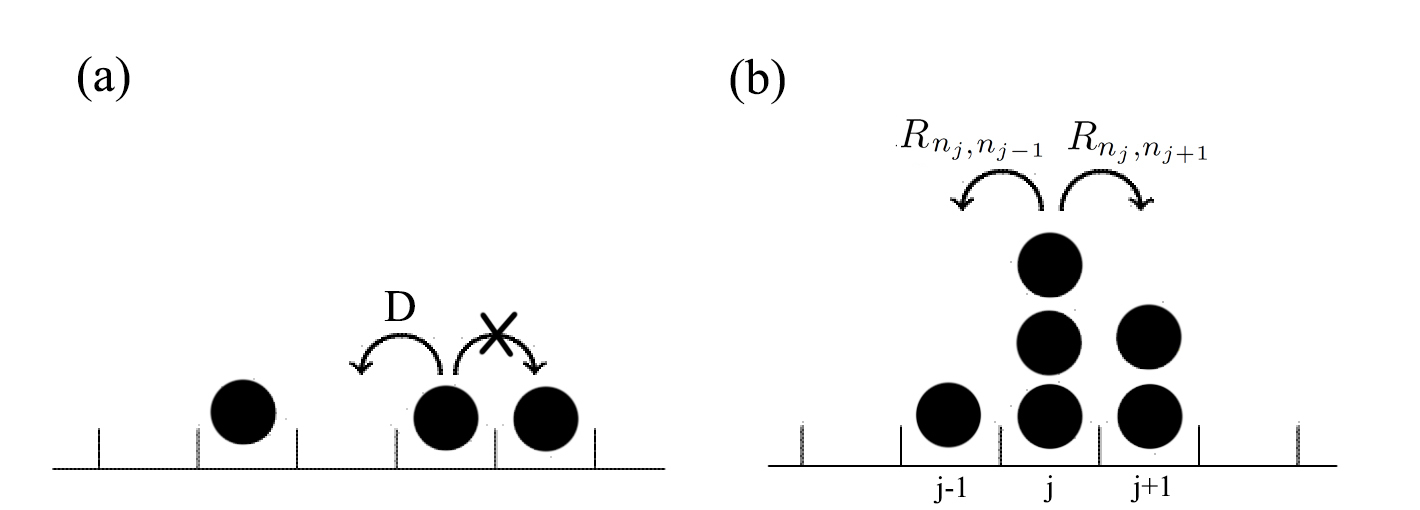}
\caption{The rates of the emerging dynamics. (a) The SSEP dynamics, where particles can jump to  empty neighboring sites with rate $D$. (b) The misanthrope model dynamics, where a particle jumps from site $j$ to site $j\pm 1$ with rate $R_{n_j,n_{j\pm 1}}$  depending on the local occupancy $n_j,n_{j\pm1}$ of site $j$ and the target site $j\pm 1 $.
\label{fig:Illustration-of-SSEP}}
\end{centering}
\end{figure}

If we interpret the local $\mathbb{P}^\pm _j$ states as site $j$ being occupied or not, the process describes the SSEP. Namely, a particle at site $j$ can hop to site $j\pm1$ with rate $D$, only if site $j\pm1$ is empty (see Fig.~\ref{fig:Illustration-of-SSEP}). For the SSEP,  the resulting dynamics are diffusive. This is starkly different than the limit of $\eta \rightarrow 0$, where the dynamics are expected to be ballistic \cite{Giamarchi_book}. 

So far we have considered a periodic chain, avoiding discussion of the boundaries. Of particular interests are the possible couplings to reservoirs, pushing the system out-of-equilibrium. 
In \footnote{The scaling $ \frac{d}{dt} \rho = L(\rho)+ \eta L_b(\rho) + \eta^{-1} L_{\textnormal{bdry}(\rho)} $ keeps the bulk and boundary dynamics on equal footing.  Then, the evolution equation is given by $\frac{d}{ds}\rho = \Pi_0 L_{\textnormal{bdry}}\Pi_0  \rho - \Pi_0  L (L^\perp _b )^{-1} L \Pi_0  \rho$. }, we explain  how to put the boundary and bulk dynamics on equal footing. We can then add the boundary dissipative terms $L^{\pm} _{\textnormal{1,N}}= \sqrt{\frac{\alpha^{\pm} _\textnormal{1,N} }{2} } \sigma^{\pm} _{1,N}$ to reconstruct the  driven SSEP dynamics explored in \cite{key5__,Derrida2004} at the large dissipation limit with $N$ sites and $\alpha^{\pm} _\textnormal{1,N}$ are the incoming/outgoing rates at sites $1,N$. 

Once the connection to the classical SSEP has been made, we can use it to learn bits of information about the original quantum system and compare its behavior with or without monitoring. Let us choose a concrete set up, widely used \cite{CFGCGF1,CFGCGF2}, to put the system away from equilibrium.  We consider the quantum system on an infinite line, preparing it with a domain wall initial state.  Namely, we set the initial time density matrix to be  $\rho_\mathrm{initial}=\rho_l\otimes\mathrm\rho_r$, with $\rho_{l,r}\propto\otimes_{i\lessgtr 0}\,e^{-\mu_{l,r}\sigma_i^z}$ where $\mu_{l,r}$ are different left/right chemical potentials. Then, we let the system evolve. The asymmetry between the left/right chemical potentials produces a spin flow through the origin that we may try to characterize. In absence of monitoring this flow is ballistic. In presence of monitoring this flow is diffusive. Because this set-up leads to the SSEP model studied in \cite{ssep-wall}, we learn  from this reference that the spin current and all spin quantum correlations dies on a diffusive time scale $1/\sqrt{t}$. Therefore, in the strong monitoring limit, the asymptotic steady state supports no current and is localized with vanishing correlation length. This is in contrast with the infinite correlation length in absence of monitoring.

\subsection{Boson chain with strong local occupancy measurements
\label{subsec: boson chain}}

Consider again a periodic chain of $L$ sites, now occupied by bosons following the tight-binding Hamiltonian $H= \varepsilon\sum_j (a^\dagger _j  a_{j+1} + a _j  a^\dagger _{j+1} )$, where $a _j , a^\dagger _{j}$ are bosonic creation and annihilation operators with the canonical commutation relations  $\big[ a _j , a^\dagger _{k}\big] = \delta_{j;k} $. We consider the local measurement operators to be $N_j = \hat{n}_j= a^\dagger _j a_j$, the local occupancy operators. Similarly to \ref{subsec: spin chain }, we find that the emerging dynamics according to \eqref{eq: strong diss evo} is 
\begin{equation}
\frac{d}{ds} \bar{\rho}_s = -\frac{1}{2} D  \sum_j \left[ a _j a^\dagger _{j+1},  \left[ a^\dagger _j a _{j+1}  , \bar{\rho}_s \right]  \right]  + \textnormal{h.c.}
\end{equation}
The pointer states are $\otimes_j \mathbb{P}^{n_j} _j $, where $\mathbb{P}^{n_j} _j = \ket{n_j}\bra{n_j}$ is the Fock space of the $j$-th site. We interpret  the pointer states as configurations,  specifying the (unbounded) number of particles at each site. Therefore, by obtaining 
\begin{eqnarray}
\mathcal{L}(\mathbb{P}^{n_j} _j  \otimes \mathbb{P}^{n_{j+1}} _{j+1} ) & =  & 
R_{n_{j}+1,n_{j+1}-1}
\mathbb{P}^{n_j +1} _j  \otimes \mathbb{P}^{n_{j+1} -1} _{j+1}    - R_{n_{j},n_{j+1}}
\mathbb{P}^{n_j} _j  \otimes \mathbb{P}^{n_{j+1}} _{j+1} 
\\ \nonumber 
&& + R_{n_{j+1}+1,n_{j}-1}\mathbb{P}^{n_j -1} _j  \otimes \mathbb{P}^{n_{j+1} +
1} _{j+1}    - R_{n_{j+1},n_{j}}
\mathbb{P}^{n_j} _j  \otimes \mathbb{P}^{n_{j+1}} _{j+1}
\end{eqnarray}
with $R_{x,y}=x(y+1)$, we can fully define the Markov process (see Fig.~\ref{fig:Illustration-of-SSEP}). A particle at site $j$ can jump to a nearby site (say $j+1$) with rate $R_{n_{j},n_{j+1}}$, depending on the local occupancy of the sites $j$ and $j+1$. We thus have found that, at the large $\eta$ limit, the  emerging dynamics of the bosonic chain is the inclusion process (with $m=2$, see \cite{Grosskinsky2011}). The shift of $y$ by $1$ in $R_{x,y}$ is a consequence of the canonical commutation relations, and hence it is an echo of the well known fact, at the core of stimulated emission, that bosons have a tendency to group.

We note however that our process satisfies  detailed balance. At this point, it is worthwhile to notice that in the initial setup, left-right symmetry is conserved. Naively, we should not   expect any breaking of detailed balance. 

In what comes next, we study the Markov limit for measurement schemes of space-dependent functions of the occupancy.

\subsection{Inhomogeneous measurements}

Let us now  consider again the bosonic chain with the tight-binding Hamiltonian. However, now we wish to consider space dependent measurements with $N_j = f_j(\hat{n}_j) $, where $f_j$ are analytic functions of their arguments. Here we stress again the requirement for non-degeneracy of the $N_j$ operators. So, we assume that the pointer states remain as in \ref{subsec: boson chain}. The dynamics of  the pointer states is then given to be 
\begin{eqnarray}
\mathcal{L}(\mathbb{P}^{n_j} _j  \otimes \mathbb{P}^{n_{j+1}} _{j+1} ) & =  & 
R^{j,j+1} _{n_{j}+1,n_{j+1}-1}
\mathbb{P}^{n_j +1} _j  \otimes \mathbb{P}^{n_{j+1} -1} _{j+1}    - R^{j,j+1}  _{n_{j},n_{j+1}}
\mathbb{P}^{n_j} _j  \otimes \mathbb{P}^{n_{j+1}} _{j+1} 
\\ \nonumber 
&& + R^{j+1,j} _{n_{j+1}+1,n_{j}-1}\mathbb{P}^{n_j -1} _j  \otimes \mathbb{P}^{n_{j+1} +
1} _{j+1}    - R^{j+1,j} _{n_{j+1},n_{j}}
\mathbb{P}^{n_j} _j  \otimes \mathbb{P}^{n_{j+1}} _{j+1},
\end{eqnarray}
where now 
\begin{equation}
R^{j,j+1}_{x,y} = \frac{x(y+1)}{\left(f_j(x)-f_j(x-1)\right)^2 + \left(f_{j+1}(y)-f_{j+1}(y+1)\right)^2 }
\end{equation}
denotes the rate of a particle jumping from site $j \rightarrow j+1$ (see Fig.~\ref{fig:Illustration-of-SSEP}). This process is known as (a subclass) of the misanthrope model \cite{key7}, the jump rate between neighboring sites $j,k$ depends only on the occupancy at the sites $j,k$. Unsurprisingly, detailed balance persists even when we carry out a space-dependent measurement scheme. While we did eliminate translational invariance symmetry, we did not explicitly break the  right-left symmetry. Thus, no current is generated and detailed balance can be recovered. 

This last fact is general: The Markov chain obtained by strongly monitoring a quantum system is always double stochastic~\footnote{Double stochasticity means that the unit vector $(1,1,\cdots)$ is both a left and right eigenvector of the Markov matrix.} as long as the system dynamics in absence of monitoring is unitary, i.e. the system Lindbladian $L$ does not contain dissipative terms. Indeed if $L(\rho)=-\frac{i}{\hbar}[H,\rho]$, then the effective Lindbladian (\ref{eq: strong diss evo}) annihilated the identity matrix, $\mathcal{L}(\mathbb{1})=0$. Thus, no current may be generated by monitoring, without feedback, a Hamiltonian system even by playing with the measurement operators (e.g. even if those operators break the left-right symmetry). 

\section{Discussion
\label{sec:Discussion}}

The behavior of lattice systems subject to strong monitoring was studied here for a few models.  The transport behavior was found to be diffusive, contrary to the ballistic transport found when no monitoring is allowed for the same models. In the strong measurement limit, the effective dynamics is that of a classical Markovian chain for which detailed balance holds, so that, as expected, no current can be generated by passively monitoring. However, since monitoring gives access to extra information via the output signals, one may choose to feedback on the system~\cite{q-feedback}, or to modulate the measurement process as in \cite{BBT-control,feed-control}, to 
e.g. break detailed balance or to make a system state follow a prescribed trajectory \cite{Mirrahimi}. 
Moreover, we have shown that at the strong measurement limit the models in question where found to follow the SSEP for fermionic chains and the inclusion process (misanthrope process) for bosonic chains. It would be interesting to see whether models with more conserved quantities will follow the nonlinear fluctuating hydrodynamics theory \cite{Spohn2015,Popkov2015}.  It may allow to explain which type of transport behavior we will encounter. 

The emerging dynamics at the large dissipation limit can be interpreted by a completely classical models, e.g. the SSEP and the inclusion process. We therefore  find  it interesting that a echo of the quantum statistics remains, i.e. fermions have an exclusion in the emerging dynamics, while bosons do not and may show a tendency to bunch. 

While the bosonic models may show a preference to bunch, they do not  condensate \cite{Grosskinsky2011,key7}. This is unsurprising, as we have an $1D$ model for bosons. Common wisdom suggests to explore a $3D$ model to be able to observe a condensation. It would be interesting to check whether for a generalization of our bosonic models, a condensation transition occurs, in the large dissipation limit and in general.

\begin{acknowledgments}
This work has been supported by ANR contract ANR-14-CE25-0003. OS would like to thank Ori Hirschberg and Takahiro Nemoto for useful discussions. DB thanks Michel Bauer for discussions and  collaborations. 
\end{acknowledgments}

% \clearpage

% a small hack to start the appendix leveled
\newpage
\begin{widetext}
\end{widetext}

\bibliographystyle{apsrev4-1}   % the bibliography style; no article names
\bibliography{refs} % the bibliography file

\end{document}